\documentclass[AMA,LATO1COL]{WileyNJD-v2} 

\articletype{Research Article}%

\usepackage{amssymb,amsfonts}
\usepackage{moreverb}
\usepackage{subfig}
\usepackage{siunitx}
\usepackage{amsmath}
\usepackage{mathtools}

\newcommand{\scolumnwidth}{8.5cm}
\newcommand{\dcolumnwidth}{17.5cm}

\begin{document}

\title{Closed-loop model-predictive wind farm flow control under time-varying inflow using FLORIDyn}

\author[1]{Marcus Becker${}^\text{1}$}

\author[2]{Maarten J. van den Broek${}^\text{2}$}
\author[3]{Dries Allaerts${}^{\text{3,}\dagger}$}
\author[1]{Jan-Willem van Wingerden${}^\text{1}$}

\authormark{BECKER \textsc{et al.}}
\titlemark{Closed-loop model-predictive wind farm flow control under time-varying inflow using FLORIDyn}

\address[1]{\orgdiv{Delft Centre for Systems and Control}, \orgname{TU Delft}, \orgaddress{\state{Delft}, \country{Netherlands}}}
\address[2]{\orgname{sowento GmbH}, \orgaddress{\state{Stuttgart}, \country{Germany}}}

\address[3]{\orgdiv{Faculty of Aerospace Engineering}, \orgname{TU Delft}, \orgaddress{\state{Delft}, \country{Netherlands}}}

\address{${}^\dagger$Deceased}
\corres{Marcus Becker. \email{marcus.becker@tudelft.nl}}

\abstract[Abstract]{
Wind farm flow control has been a key research focus in recent years, driven by the idea that a collectively operating wind farm can outperform individually controlled turbines.
Control strategies are predominantly applied in an open-loop manner, where the current flow conditions are used to look up precomputed steady-state set points.
Closed-loop control approaches, on the other hand, take measurements from the farm into account and optimize their set points online, which makes them more flexible and resilient.

This paper introduces a closed-loop model-predictive wind farm controller using the dynamic engineering model FLORIDyn to maximize the energy generated by a ten-turbine wind farm.
The framework consists of an Ensemble Kalman Filter to continuously correct the flow field estimate, as well as a novel optimization strategy.
To this end the paper discusses two dynamic ways to maximize the farm energy and compares this to the current look-up table industry standard.
The framework relies solely on turbine measurements without using a flow field preview.
In a 3-hour case study with time-varying conditions, the derived controllers achieve an overall energy gain of $3$ to $4.4$~\% with noise-free wind direction measurements.
If disturbed and biased measurements are used, this performance decreases to $1.9$ to $3$~\% over the greedy control baseline with the same measurements.
The comparison to look-up table controllers shows that the closed-loop framework performance is more robust to disturbed measurements but can only match the performance in noise-free conditions.
}


\maketitle

\section{Introduction}\label{sec1: Introduction}
A switch away from fossil fuels to less greenhouse gas-emitting (GHG) sources of energy is necessary to prevent a climate crisis \citep{uneceCarbonNeutralityUNECE2022}. Wind energy is one alternative that provides energy at a fraction of the GHG emissions.
Wind farms are, therefore, an essential part of the energy transition. However, they do not provide the maximum amount of energy they could. How come? This is in part due to the way that turbines interact: As a wind turbine converts kinetic energy from the surrounding airflow into electricity, it leaves behind an area of low wind speed called a wake. In wind farms, these wakes will likely influence downstream turbines, lowering their power output. Wind farm flow control (WFFC) strategies aim to mitigate this effect by manipulating the wake shape. 

In this work, we focus on model-based approaches to WFFC. With this approach, a model of the farm is used as a surrogate for the real wind farm. The model predicts how the turbine wakes behave, given the atmospheric conditions and turbine states.
Different types of models exist with varying costs and capabilities. On the one end, high-fidelity models like Large Eddy Simulations (LES) provide the most insight into the physical phenomena that take place in a wind farm, e.g. \cite{churchfieldNumericalStudyEffects2012, chatelainLargeEddySimulation2013}. These models are typically too demanding for control which has motivated model simplifications to achieve faster computational speeds. Medium-fidelity models capture a coarse image of the flow. Two-dimensional Reynolds-averaged Navier–Stokes solvers like e.g., \cite{rottDynamicFlowModel2017, boersmaControlorientedDynamicWind2018, vandenbroekDynamicFlowModelling2020} fall in this category. They aim to capture the core wake dynamics at a reduced cost by limiting the dimensionality of the flow. However, later research showed that this simplification can render them unuseful for wake steering applications \citep{vandenbroekFlowModellingWind2022}. Other models in the category of medium-fidelity models simulate the wake propagation based on synthetic turbulence. The Dynamic Wake Meandering (DWM) model, introduced by \cite{larsenDynamicWakeMeandering2007}, propagates the wake as a series of turbulence boxes. The propagation speed and direction are then determined by the contents of the box. This approach, coupled with an aeroelastic turbine model, can give estimates of loads onto the turbine and its structure. Successor models like FAST.Farm \citep{jonkmanDevelopmentFASTFarm2017} and HAWC2Farm \citep{liewDynamicModellingWind2022} provide a basis to investigate damage equivalent loads on a farm scale at a relatively low computational cost.
Another approach to model wind turbine wakes is to simulate free-vortex particles shed by the rotor, e.g., \cite{martenQBladeModernTool2020, vandenbroekAdjointOptimisationWind2022}. The downside of these models is that they tend to become numerically unstable in the far-wake region and under turbulent conditions.
Low-fidelity models provide a flow field prediction at a very low computational cost. This is achieved by modeling the wake shape and wind speed reduction as a set of analytical equations, e.g., \cite{bastankhahExperimentalTheoreticalStudy2016}. These models typically predict the steady-state wake shape of a single turbine wake and use superposition methods to combine the effect of multiple wakes in a farm. Within WFFC applications, they are used to test and optimize yaw-angle set points for an entire farm. 

Based on the computational speed provided by the steady-state engineering models, paired with the success of the DWM model, an additional group of models has been proposed: These models use passive Lagrangian tracers, called Observation Points, to propagate turbine and flow field states from each turbine downstream. The Flow Redirection and induction Dynamics model FLORIDyn initially proposed this approach \citep{gebraadControlOrientedDynamicModel2014}. Since its introduction, the model has since been revised \citep{beckerRevisedFLORIDynModel2022a} and further developed \citep{beckerFLORIDynDynamicFlexible2022a}. Similar modeling approaches exist, e.g., \cite{lejeuneMeanderingCapturingWakeModel2022, braunbehrensApplicationOpenloopDynamic2022}.
Their simplicity, simulated wake dynamics, and speed make them attractive for model-based WFFC applications. 

\begin{figure*}
    \centering
    \includegraphics[width=\linewidth]{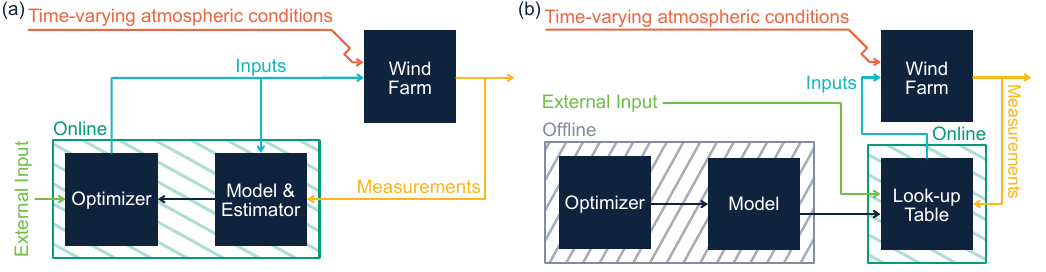}
    \caption{(a) Closed-loop compared to (b) open-loop wind farm flow control. The figure is adapted from \cite{meyersWindFarmFlow2022}.}
    \label{fig: open-loop vs closed-loop}
\end{figure*}
Model-based WFFC is predominantly applied in an open-loop configuration \citep{meyersWindFarmFlow2022}. This means that the yaw angle set points are optimized ahead of time; see Figure \ref{fig: open-loop vs closed-loop} (b). This is mainly done by employing a steady-state wake model and selecting a set of ambient conditions. For each ambient condition, the yaw angle set points are then optimized and stored in a Look-up Table (LuT). During operation, the current ambient conditions, such as wind speed and direction, are identified and used to look up the optimal yaw angles. Examples of this strategy can be found in the field experiments conducted by e.g., \cite{flemingFieldTestWake2017, ducLocalTurbulenceParameterization2019, doekemeijerFieldExperimentOpenloop2021}.
The issue with this approach is that the strategy can not react to unforeseen circumstances. 
Examples of this could be wind farm layout changes due to offline turbines, unmodeled or incorrectly modeled turbine interactions, or heterogeneous and changing flow conditions.

This is addressed by closing the loop, see Figure \ref{fig: open-loop vs closed-loop} (a), where the model is continuously updated by measurements.
A recent approach is to close the loop on the parameters of the steady-state model \citep{doekemeijerClosedloopModelbasedWind2020a, howlandOptimalClosedloopWake2020, howlandOptimalClosedloopWake2022, soodDevelopmentValidationLarge2023}. This entails a coupling between the measured flow conditions and the modeled ones. The difference is then fed back to correct aspects like the wake expansion. In \cite{bachantDevelopmentValidationHybrid2024a}, they employ a similar strategy, but rather than correcting the model parameters, they build a corrector for the model output. This version of closed-loop control still inherently neglects the wake dynamics. 
Closed-loop wind farm flow control using a dynamic wake model like FLORIDyn is still a poorly explored area of research. \cite{gebraadWindTurbineWake2015a} employs the first version of the FLORIDyn model paired with a Kalman Filter to estimate the wind speed in a six-turbine wind farm. The same publication also uses FLORIDyn to perform a yaw angle optimization; however, not in closed-loop. 

Control for energy maximization using dynamic models uses predominantly a receding horizon approach called Model-predictive control (MPC), e.g., \cite{goitOptimalControlEnergy2015, muntersDynamicStrategiesYaw2018, vandenbroekAdjointOptimisationWind2022, beckerTimeshiftedCostFunction2024}. This means that the optimization is done over a predetermined prediction horizon. The surrogate model is used to predict how the cost function will be impacted by the control actions taken. An optimizer is then used to determine the ideal control actions. The optimization is followed by an update time step, which is typically smaller than the prediction horizon. During this time, the previously optimized time series of control set points is applied. After the update time step has passed, a new optimization is done. MPC is typically applied to follow a reference value, e.g. a reference wind farm power \citep{sterleModelPredictiveControl2024}. In contrast this work aims to maximize a cost function, which leads to a different optimization problem. This is referred to as economic MPC (eMPC).

A re-occurring assumption in this context is the full knowledge of the flow field (e.g. \cite{goitOptimalControlEnergy2015, goitOptimalCoordinatedControl2016, muntersDynamicStrategiesYaw2018, janssensRealtimeOptimalControl2024}), and full knowledge of the environmental changes ahead (e.g., \cite{vandenbroekAdjointOptimisationWind2022, vandenbroekDynamicWindFarm2024}). The latter is often referred to as preview and can lead to significant gains over preview-less control approaches, also for steady-state control approaches \citep{simleyWakeSteeringWind2021, sengersIncreasedPowerGains2023a, vandenbroekDynamicWindFarm2024}. Yet, it is not clear how this preview information may be attained in a realistic WFFC scenario. The same holds true for the previously mentioned assumption of full flow field knowledge.
State-estimation techniques can provide some of this knowledge. They use sensor data to correct the current state of the surrogate model. This is done by comparing predicted measurements to taken ones. Looking at FLORIDyn specifically, in the work by \cite{gebraadWindTurbineWake2015a}, the turbine power generated is used to correct the local wind speed estimate in FLORIDyn by employing a Kalman Filter. Similarly, \cite{beckerEnsembleBasedFlowField2022} also uses the power generated, as well as wind direction measurements, to correct a heterogeneous flow field state in FLORIDyn using an Ensemble Kalman Filter (EnKF). This methodology has the advantage that it provides the state estimate as well as an uncertainty estimate. Additionally, it does not require a linear(-ization) of the model. In the later work of \cite{braunbehrensMultiscaleKalmanFiltering2023} and \cite{dicaveClosedloopCouplingDynamic2024}, the wake location is corrected by using an EnKF and a downstream turbine as a coarse sensor~\citep{schreiberFieldTestingLocal2020}.

\begin{figure*}
    \centering
    \includegraphics[width=1\linewidth]{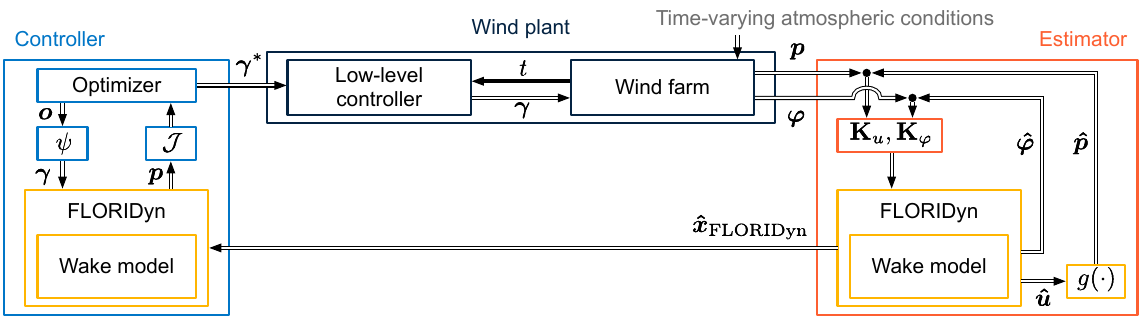}
    \caption{Closed-loop design applied in this paper. The wind farm provides power and wind direction measurements at the turbine locations, which are used to correct the FLORIDyn model state. The identified state is then used to optimize the future yaw angle set points for all turbines. This is passed on to a low-level controller, which applies the set points.}
    \label{fig: closed-loop design}
\end{figure*}

The literature review shows that there is a gap in closed-loop wind farm flow control approaches using dynamic surrogate models in time-varying conditions.
Prior work has developed surrogate models like FLORIDyn to dynamically model the wakes in a farm. It has also proposed state-estimation frameworks to align the model’s state with the true state. How to then use FLORIDyn to obtain the optimal control set points is still an unexplored topic.

This paper, therefore, proposes a novel closed-loop economic model predictive control framework for farm-wide energy maximization. To this end, the paper builds upon previous work with the following four main contributions:
(i) a discussion around the nature of the cost functions for energy and power maximization, (ii) a dimensionality reduction of the optimization problem, which also leads to a realistic turbine operation, (iii) an implicit way to limit misalignment angles and to avoid nonlinear constraints, and (iv) the derivation of a closed-loop framework for dynamic model-based wind farm flow control. The proposed framework focuses on readily available sensor data, like turbine power and wind direction measurements at the turbine locations, and in contrast to previous work, does not assume knowledge of the full flow field nor a preview of future flow field changes.
The derived optimization strategy is applied in a ten-turbine closed-loop case study. This is done with a wind farm simulated in LES under turbulent conditions and with realistic time-varying wind direction changes. In this context, the paper also addresses the impact of noisy and biased measurements on controller performance and yaw travel.

The remainder of the paper is structured as follows: Section \ref{sec: Methodology} discusses the methodology of the closed-loop approach. This consists of a description of the model (Section \ref{sec: meth: FLORIDyn}), the state estimator (Section \ref{sec: meth: state estimation}), and the novel discussion related to the controller (Section \ref{sec: meth: controller}). 
The simulation methods are discussed in Section \ref{sec: Sim methods}. This entails the description of the high-fidelity environment, as well as the tested wind conditions, sensor data, and controller settings.
The proposed framework is tested in Section \ref{sec: Results: simulation results}, where the results are presented. Lastly, Section \ref{sec: Conclusion} draws a conclusion of the work.

\section{Methodology}\label{sec: Methodology}
This section introduces the components of the closed-loop control approach presented in this paper: Section \ref{sec: meth: FLORIDyn} describes the basics of the dynamic surrogate model, Section \ref{sec: meth: state estimation} of the state estimation. The main contribution of this paper, the controller, is discussed in Section \ref{sec: meth: controller}. The list of parameters used for the components introduced in Section \ref{sec: Methodology} can be found in Appendix \ref{app: parameters}, alongside the tuning approach.

\subsection{Surrogate model}\label{sec: meth: FLORIDyn}
The Flow Redirection and Induction Dynamics Model (FLORIDyn) used in this study is based on the work presented in \cite{beckerFLORIDynDynamicFlexible2022a}, with the extensions done in \cite{beckerEnsembleBasedFlowField2022}.
It simulates wake dynamics by employing so-called Observation Points (OPs), which carry states from the rotor plane downstream. The states one OP possesses are (i) turbine states (e.g., yaw angle), (ii) flow field states (e.g., wind direction), and (iii) its own positional states.
The states are initialized at the rotor plane based on what the turbine measures. The OP position is advanced based on the wind speed and direction at its position.

While previously each OP would rely on its own state of the wind speed and direction to propagate, the wind speed and direction are now the result of a weighted average. This way, the OP takes neighboring OP states into account. The spatiotemporal Gaussian weighting function was first adapted by \cite{lejeuneMeanderingCapturingWakeModel2022} for a FLORIDyn similar model and adapted in \cite{beckerEnsembleBasedFlowField2022} to allow for better correlation between the OPs. This benefits the assumptions made with the state estimation; see Section \ref{sec: meth: state estimation}.
The FLORIDyn framework uses a Gaussian wake to model the wake deficit, wind speed reduction and wake deflection  \citep{bastankhahExperimentalTheoreticalStudy2016}.

Unlike recently presented results (e.g. \cite{vandenbroekDynamicWindFarm2024}) the simulations presented in this paper do not utilize any synthetic preview information of the wind direction. As previously mentioned, FLORIDyn does have flow field information stored in the OPs, which is propagated downstream. This provides other turbines with an estimate of the flow field changes ahead. During the prediction phase, each time step, the free stream turbines copy their previous state, and downstream turbines adapt the wind speed and direction by the weighted average of the data stored in the surrounding OPs.

\subsection{State estimation}\label{sec: meth: state estimation}
\begin{figure}
    \centering
    \includegraphics[width=\scolumnwidth]{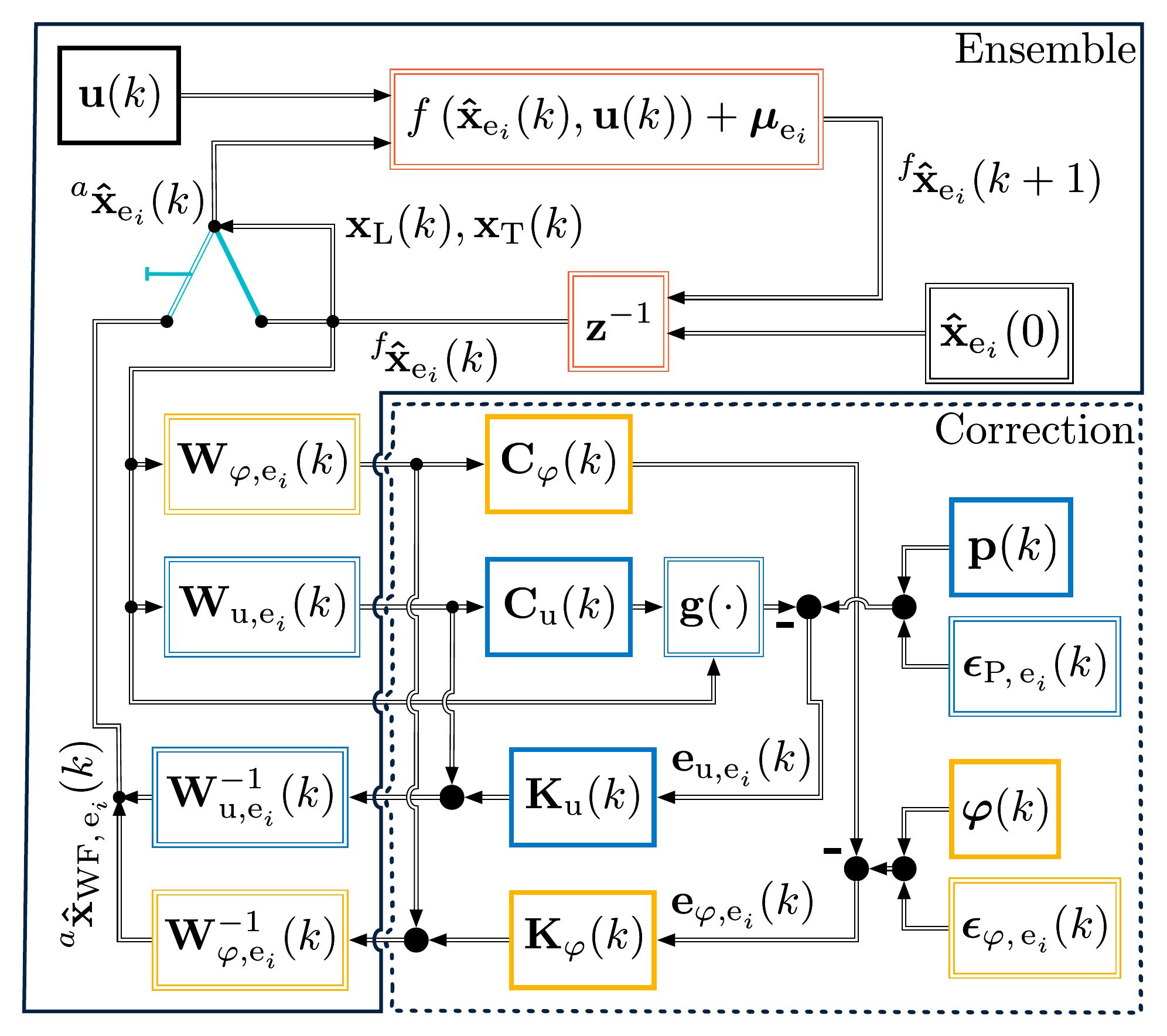}
    \caption{State estimation cycle for an ensemble $e_i$. The upper section of the figure depicts the evolution of the wind farm simulation, here depicted by $f(\cdot)$, the lower left section the extraction of the measured values from the simulation. The lower right part depicts the correction based on the mismatch of the predicted and recorded measurements. Elements with a double outline are ensemble-specific, elements with a single outline are the same for all ensembles.}
    \label{fig: enkf}
\end{figure}

The employed state estimation is based on \cite{beckerEnsembleBasedFlowField2022} and uses an Ensemble Kalman Filter (EnKF) \citep{evensenSequentialDataAssimilation1994, evensenDataAssimilationEnsemble2009}. An EnKF was already previously used by \cite{doekemeijerEnsembleKalmanFiltering2017} to estimate the flow field state of a wind farm simulation in the 2D solver WFSim \citep{boersmaControlorientedDynamicWind2018}. What distinguishes a dynamic system like FLORIDyn from this application is that FLORIDyn is not a grid-based simulation but rather attaches its state to particles. Therefore, the flow is described by where the particles are, which may be different across the ensembles.
The state estimation framework used addresses this issue by projecting the OPs of all Ensembles onto a common set, which is then corrected. To this end, two Kalman gain matrices ($\mathbf{K}_u$ for wind speed and $\mathbf{K}_\varphi$ for the wind direction) are derived from the common output matrices $\mathbf{C}_u$ and $\mathbf{C}_\varphi$, as well as the correlation of the ensembles. 
This is done differently to the state estimator proposed in \cite{beckerEnsembleBasedFlowField2022}, where only the wind speed would be corrected this way, while the wind direction would be corrected with ensemble individual Kalman Gain matrices.
The correction loop is depicted in Figure \ref{fig: enkf}. The matrices $\mathbf{W}$ define the weighted projection onto the common states, $\mathbf{W}^{-1}$ their inverse. Since the inverse is generally not attainable, we set it to be equal to the identity matrix. Since $\mathbf{W}$ does have a sparse and strongly diagonal shape, this approximation is valid.
For more information on how the correction is derived we refer to the previously mentioned sources.
There are EnKF-based designs for FLORIDyn-like models to correct the wake center, but these fall outside of the scope of this paper \citep{braunbehrensMultiscaleKalmanFiltering2023, dicaveClosedloopCouplingDynamic2024}. Also outside of the scope of this paper fall online parameter estimation methods using an EnKF \citep{howlandOptimalClosedloopWake2020, doekemeijerClosedloopModelbasedWind2020}.

\subsection{Controller}\label{sec: meth: controller} 
Section \ref{sec: meth: cost function} compares the differences between the steady-state cost function formulation and two dynamic cost functions.
Section \ref{sec: meth: yaw angle derivation} to \ref{sec: meth: wind farm decomposition} discuss measures to simplify the optimization problem solved at runtime:
Section \ref{sec: meth: yaw angle derivation} investigates a basis function approach to derive the yaw angle time series, followed by a methodology to incorporate constraints into the cost function (Section \ref{sec: meth: yaw limitation}). 
Section \ref{sec: meth: wind farm decomposition} discusses how the wind farm is decomposed into smaller sections to reduce the number of turbines per optimization. Lastly, Section \ref{sec: meth: reference controllers} discusses the reference controllers.

\subsubsection{Cost function}\label{sec: meth: cost function}
\begin{figure}
    \centering
    \includegraphics[width=\scolumnwidth]{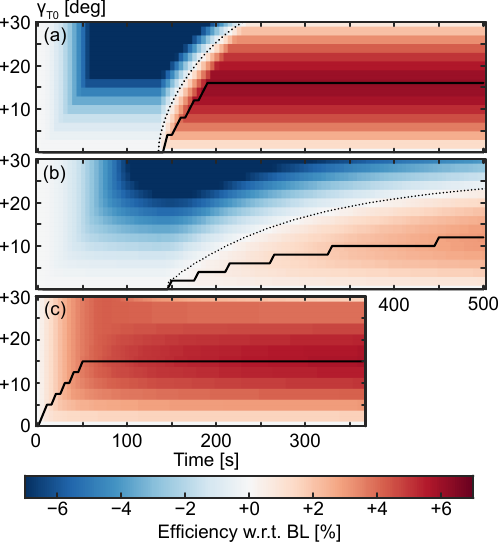}
    \caption{Wind farm power (a), energy (b), and shifted energy (c) efficiency of a two turbine wind farm based on the yaw control set point of T0, normalized by the case of no misalignment. Turbine T0 starts at $t=0$ s with $\gamma_\text{T0} = 0$ deg and continuously changes its angle with $0.3 \text{ deg s}^{-1}$ to the set point, where it remains. Turbine T1 is fully aligned with the wind direction. The dotted line shows the break-even line, the continuous line the ideal value of $\gamma_\text{T0}$ based on the time. Since the shifted approach (c) moves the power signal by T1 in time, the data is shorter.}
    \label{fig:two_turbine_experiment}
\end{figure}

The steady-state approach to formulate a cost function $\mathcal{J}_\text{steady state}$ to maximize the farm energy is to sum the power of all $n_T$ turbines:
\begin{equation}
    \mathcal{J}_\text{steady state}(\theta) = \sum_{i=1}^{n_T} p_i(\theta)\,,\label{eq:con:ss_cost_func}
\end{equation}
where $p_i$ is the power generated by turbine $i$ and $\theta$ are the optimization variables. Time is implicitly taken into account, as steady-state models do not assume any delay effects. If time is taken into account, the energy is calculated as the integral of the turbine power generated over time. Equation \eqref{eq:con:ss_cost_func} is extended as follows to form the dynamic cost function:
\begin{equation}
    \mathcal{J}_\text{dynamic}(\theta) = \Delta t\sum_{i=1}^{n_T} \sum_{k=1}^{\tau_\text{ph}}p_i(\theta,k)\,,\label{eq:max_energy}
\end{equation}
where $p_i(\theta, k)$ is the power generated by turbine $i$ at time step $k$, and $\Delta t$ is the time step of the simulation. The cost function now approximates the energy generated by the farm across all time steps $k\in[1,\tau_\text{ph}]$, where $\tau_\text{ph}$ is the prediction horizon of the optimization. This additional time dependence inherently changes the result for $\theta$ that we get: The power loss due to yaw control actuation needs to be recouped within the prediction horizon for the control action to be profitable. This is in strong contrast to the steady-state cost function, which assumes to have eternity available to recoup the cost of control actions.\\
The following example further illustrates this difference based on a two-turbine wind farm: turbine T1 is located $5D$ downstream and $0.5D$ cross stream from the upwind turbine T0, $D$ being the turbine diameter. This configuration favors positive yaw misalignment from the first turbine for power and energy maximization. In this experiment, the control set point of T0 is varied, and the power generated from both turbines is recorded in FLORIDyn. At $t=0$ s, T0 starts fully aligned with the wind direction and then changes its orientation with $0.3 \text{ deg s}^{-1}$ to the reference misalignment angle. The chosen yaw speed is within the commonly used range with up to $0.5\text{ deg s}^{-1}$ \citep{kimYawSystemsWind2014}.

Figure \ref{fig:two_turbine_experiment} shows the efficiency of the yaw misalignment angle set point over time, (a) based on the power generated, and (b) based on the energy generated up to the given time step. Both figures show the initial loss connected to the misalignment of turbine T0. Turbine T1 starts to experience the benefits of the misalignment after $\approx 150$ to $200$ s. At this point, the steady state optimum is reached for $\gamma_\text{T0}\approx +16 $ deg, which leads to a $7\,\%$ increase in power generated compared to the baseline case. The energy generated, however, has integrated the losses leading up to the favorable wake redirection state. This means that this loss has to be recouped first before a yaw misalignment becomes a better choice than the baseline behavior. The more time is given, the more attractive larger yaw angles become: If the prediction horizon is $100$ s long, $\gamma_\text{T0}=0 $ deg is the optimal solution to Equation \eqref{eq:max_energy}, for $300$ s it is $\gamma_\text{T0}\approx8 $ deg and for $500$ s $\gamma_\text{T0}\approx12 $ deg is optimal. With an infinite prediction horizon, the optimal solution of Equation \eqref{eq:max_energy} converges to the optimal solution of Equation \eqref{eq:con:ss_cost_func}, as the initial loss becomes increasingly negligible compared to the accumulated energy. This inherent characteristic of Equation \eqref{eq:max_energy} offers a conservative security, as it guarantees to recoup the initial investment in the given time frame under the assumption that conditions are steady. From this perspective wake steering becomes an investment-and-return problem, threatened by changing environmental conditions. 

The described characteristics can also lead to unwanted behavior - if enough control degrees of freedom are given, T0 will initially engage in more aggressive control actions, followed by greedy behavior at the end of the time horizon. This is done since the downsides of the greedy control actions towards T1 fall outside of the time horizon. This is known as the "turnpike effect" and needs to be considered for dynamic wind farm flow control applications \citep{vandenbroekAdjointOptimisationWind2022}. Ways to limit the turnpike effect is to not allow control actions at the end of the time horizon or to penalize them as part of the cost function.

An alternative is offered by \cite{beckerTimeshiftedCostFunction2024}: The presented approach aims to synchronize the cause and effect of the control action with its impact on the actuated turbine as well as the downstream turbines. This formulation leads to a spatiotemporal split of the optimization problem into several smaller problems: (i) the spatial split ensures that wind turbines that do not influence one another are optimized separately, while (ii) the temporal split connects the cost function of the yaw angles with their impact.
The following example illustrates the working of the algorithm using the same two-turbine wind farm. The free wind speed is set to $8 \text{ms}^{-1}$, and the wake advection speed is assumed to be equal to the free stream velocity. The simulation time step is $\Delta t = 5$~s. It then takes $800\text{ m}/8\text{ ms}^{-1} = 100$ s or $\Delta k = 20 $ time steps for one particle of air to reach the downstream turbine. If the action horizon is set to $\tau_\text{ah} = 10$ time steps, the cost function of the yaw trajectory $\boldsymbol{\gamma}_\text{T0}$ of the upstream turbine T0 is formulated as follows:
\begin{align*}
    \min\mathcal{J}_\text{T0}(\boldsymbol{\gamma}_\text{T0}) = -\Delta t \sum_{k=1}^{\tau_\text{ah}}\, &p_0(\boldsymbol{\gamma}_\text{T0},k) + 
    p_1(\boldsymbol{\gamma}_\text{T0}, \boldsymbol{\gamma}_\text{T1}, k + \Delta k)\,.
\end{align*}
This formulation sums the power of T0 during its action horizon and the power of T1 once the control actions arrive at the turbine.
The power of T1 depends on both $\boldsymbol{\gamma}_\text{T0}$ and $\boldsymbol{\gamma}_\text{T1}$. Since $\Delta k_{1\leftarrow0} > \tau_\text{ah}$, the yaw trajectory $\boldsymbol{\gamma}_\text{T1}$ can be optimized in a separate cost function:
\begin{align*}
    \min\mathcal{J}_\text{T1}(\boldsymbol{\gamma}_\text{T1}) = -\Delta t \sum_{k=1}^{\tau_\text{ah}}\, &p_1(\boldsymbol{\gamma}_\text{T1},k)\,.
\end{align*}
As a result we can formulate two optimization problems, $\mathcal{O}_1 = \min\mathcal{J}_\text{T0}$ and $\mathcal{O}_2 = \min\mathcal{J}_\text{T1}$. In this example $\mathcal{O}_1$ depends on $\boldsymbol{\gamma}_\text{T1}$, and thus on the result of $\mathcal{O}_2$. Therefore, $\mathcal{O}_1$ cannot be solved independently. $\mathcal{O}_2$ is only evaluated over $\tau_\text{ah}$ time steps, and $\mathcal{O}_1$ over $\tau_\text{ah} + \Delta k_{1\leftarrow0} $ time steps, which is generally less than the otherwise necessary $\tau_\text{ph}$ to achieve similar yaw steering. Additionally, the smaller optimization problems can lead to better convergence within the given optimization budget. More details about the time-shifted approach can be found in \cite{beckerTimeshiftedCostFunction2024} or in a similar application in \cite{ciriModelfreeControlWind2017}. Figure \ref{fig:two_turbine_experiment} c) shows how the optimization landscape is affected by the change.

\subsubsection{Basis function approach}\label{sec: meth: yaw angle derivation}
\begin{figure}
    \centering
    \includegraphics[width=\scolumnwidth]{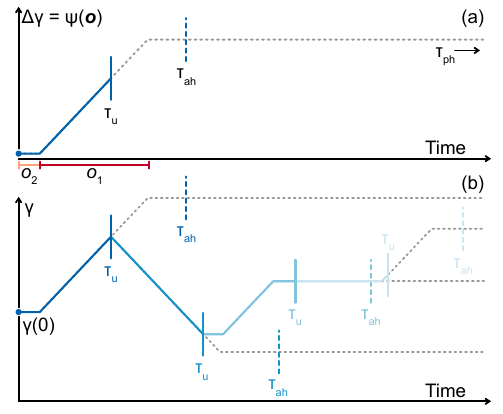}
    \caption{Example of $\psi$ and the resulting displacement $\Delta\gamma$ from the current orientation (a). The percentage of the action horizon $\tau_\text{ah}$ that is spent yawing is determined by $o_1$, while $o_2$ determines the offset at which the angle change starts. At $\tau_\text{u}$, the optimization is redone based on the updated model state. An example of how a turbine orientation might change over successive optimizations is given in (b).}
    \label{fig:2dofmr}
\end{figure}

The optimized yaw time series must fulfill two constraints: (i) its rate of change needs to be lower or equal to the possible rate of change~$r_\gamma$, limited by the actuators, and (ii) it should not exceed the prescribed maximum misalignment boundaries. This section focuses on the former aspect, while the latter is discussed in Section \ref{sec: meth: yaw limitation}. In addition to the limited rate of change, it is desirable to derive yaw signals that can be realistically applied to existing yaw systems. Current yaw systems are comprised of a yaw drive, a bearing, and a brake \citep{kimYawSystemsWind2014}. To move the turbine, the brake is released and the yaw drive moves the nacelle with the maximum rate of change to the reference position. The brake is then engaged again to maintain the reference orientation. This behavior limits the set of feasible yaw time series, as, e.g., a continuously changing yaw signal is undesirable. This can be exploited to simplify the optimization problem and to reduce the number of inputs to the optimization. 
To this end, we rely on a basis function $\psi(\boldsymbol{\theta},t_n)$, which takes two optimization parameters, $o_1$ and $o_2$ as arguments, as well as $t_n\in [0,1]$, which is the normalized time within the action horizon $t_\text{AH}$. The idea is to either in- or decrease the turbine's yaw angle with the maximum rate of change and to allow the change period to start at an early or late point in time within $t_\text{AH}$. The function $\psi$ has been designed in such a way that $o_1,o_2 \in [0,1]$, which provides a generic interface for an optimization algorithm and can improve its effectiveness of it.
The basis function is further determined by $r_{\gamma,n}=r_\gamma\,t_\text{AH}$, the maximum rate of change within the normalized parameter space. 
\begin{align}
    \psi(\boldsymbol{o},t_n) &= 2[o_1-0.5]\,\text{sat}_{[0,1]}\left(\frac{t_n - t_{s,n}(\boldsymbol{o})}{2|o_1-0.5|}\right)\,r_{\gamma,n}\,,\label{eq:2dof_mr}\\
    t_{s,n}(\boldsymbol{o}) &= o_2\Bigl[1-2\cdot|o_1-0.5|\Bigr]\,,\nonumber
\end{align}
where $\text{sat}_{[0,1]}(x)$ is a saturation function which is equal to $0$ for $x\leq0$, $x$ for $x\in[0,1]$, and $1$ otherwise. The variable $o_1$ determines in which direction the orientation changes and for how long, e.g. $o_1=1$ is a constant increase across the entire action horizon, $o_1=0.5$ results in a steady yaw angle. The second optimization variable, $o_2$, moves the starting point of the yaw angle change period. A possible result of $\psi$ and the resulting $\Delta\gamma$ is shown in Figure \ref{fig:2dofmr}(a). 
The final yaw orientation trajectory is calculated as 
\begin{equation}
    \gamma(t) = \gamma(0) + \psi\left(\boldsymbol{o},\frac{t}{t_\text{AH}}\right) \quad \forall t\in[0,t_\text{AH}]\,.
\end{equation}
An example is given in Figure \ref{fig:2dofmr}(b).
While the proposed function does limit the number of optimization parameters per turbine, it contains undesired regions of insensitivity. This namely affects the case when $o_1=0.5$, which results in no yaw change and removes any effect of $o_2$ onto the resulting trajectory. Similarly, for $o_1=0$ or $=1$, the entire duration of $t_\text{AH}$ is spent yawing, which also nullifies the impact of $o_2$. This impacts the optimization landscape and introduces regions with no gradient sensitivity towards $o_2$. This affects the optimizer choice in Section \ref{sec:con:opti}.

\subsubsection{Implicit misalignment limitation}\label{sec: meth: yaw limitation}
The methodology introduced in Section \ref{sec: meth: yaw angle derivation} limits the possible solution space of the yaw angle trajectories based on the yaw rate limit. However, it does not enforce a limit on what the resulting yaw angle might be. 
There are different ways of achieving a yaw misalignment limit, one would be to set it as a constraint. This is a linear constraint if the yaw misalignment is the input to the optimization problem. However, if wind direction changes are present, it is beneficial to optimize the turbine orientation instead of the misalignment, as the misalignment is a product of the uncontrollable wind direction and the turbine orientation. This switch, along with the basis function approach discussed in Section \ref{sec: meth: yaw angle derivation}, leads to a nonlinear constraint of the yaw angle misalignment. This creates an additional layer to implement and can increase the complexity of the optimization problem. We instead chose to manipulate the power calculation in such a way that large yaw angles become much less attractive. This removes the constraint and makes an implicit part of the cost function. To this end, we formulate a weighting function for the power generated:
\begin{align}
    w_\gamma(\gamma, \gamma_\text{max},\gamma_\text{min}) =& \left[\frac{1}{2}\tanh(50\left[-\gamma+\gamma_\text{max}\right]) + \frac{1}{2}\right] \cdot \dots \nonumber
    \\& \left[-\frac{1}{2}\tanh(50\left[-\gamma+\gamma_\text{min}\right]) + \frac{1}{2}\right].\label{eq:con:opti:yaw limitation}
\end{align}
The weighting function reduces the power generated by $50\%$ at the limit yaw angles and smoothly transitions from the unmodified part of the power curve to the lowered part outside of the limits.
The limits were chosen as $\gamma_\text{max} =-\gamma_\text{min}=33~\text{deg}$. Note that Equation \eqref{eq:con:opti:yaw limitation} reduces the power for yaw angles that are still within but close to the bounds. As a result, yaw angles $>33$~deg are already unattractive to the optimizer.

\subsubsection{Wind farm decomposition}\label{sec: meth: wind farm decomposition}
\begin{figure}
    \centering
    \includegraphics[width=\scolumnwidth]{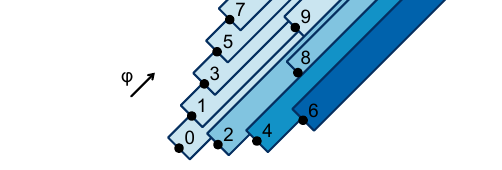}
    \caption{Demonstration of the wind farm decomposition into smaller subsets, based on the $\pm2$~D crosswind distance selection. Turbines that are grouped together share the same color.}
    \label{fig: WF decomposition}
\end{figure}
To reduce the computational cost of the optimization and to ensure a faster convergence, a decomposition of the wind farm in smaller subsets is useful \citep{bernardoniIdentificationWindTurbine2021}. This ensures that turbines that are unaffected by other turbines, nor affect any, can optimize for greedy behavior. Turbines that do affect one another are optimized together. In this work, we decompose the wind farm based on the current wind direction at each turbine's location. Each turbine calculates which other turbines that are within $\pm 2$ D crosswind distance and downstream based on the current wind direction, see Figure \ref{fig: WF decomposition}. These then become part of the optimization group related to this turbine. We then recursively determine which turbines affect one another. This creates a directed graph, similar to the model principles proposed by \cite{starkeDynamicModelWind2023}. Each group is then optimized together, meaning that only those turbines in the group are simulated. This effectively reduces the ten-turbine wind farm to smaller farms with one to six members. This wind farm decomposition is done with every optimization step given the current conditions and, therefore, changes over time.

\subsubsection{Optimization algorithm}\label{sec:con:opti}
The presented framework uses a particle swarm optimization to solve the cost function, which is part of the evolutionary algorithms \citep{ieeeneuralnetworkscouncilProceedings1995IEEE1995, pedersenGoodParametersParticle2010, mezura-montesConstrainthandlingNatureinspiredNumerical2011}. The code uses the implementation by \cite{mathworksMATLABOptimizationToolbox2023}.
The optimizer was chosen for the main reason that the optimization landscape is not convex. This stems from the split nature of wake steering, as steering in both directions might yield an improvement while one is favorable. Another contribution is the insensitivity of yaw angle basis function for certain parameter combinations, see Section \ref{sec: meth: yaw angle derivation}. 
A total of 100 particles is used for all optimizations. The stopping criterion is either a maximum number of four consecutive iterations with no cost function improvement or 20 iterations. 
These settings were chosen after also testing 2, 10 and 40 iterations. 
This design further leverages the code's capability to run multiple FLORIDyn instances in parallel, similar to the EnKF. Gradient-based methods have been tested during the development of the closed-loop controllers but have been outperformed by non-gradient-based methods. The development of a dedicated optimization strategy like the Serial Refine method \citep{flemingSerialRefineMethodFast2022} lies outside of the scope of this paper.

\subsection{Reference controllers}\label{sec: meth: reference controllers}
\begin{figure}
    \centering
    \includegraphics[width=\scolumnwidth]{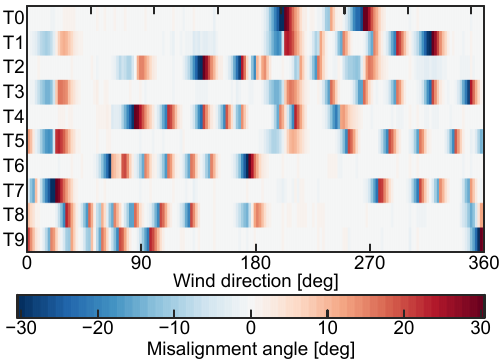}
    \caption{Look-up table generated with the internal FLORIDyn wake model for the reference controllers.}
    \label{fig: lut}
\end{figure}

We use two types of reference dead-band lookup-table (LuT) yaw steering controllers, one that aims for $0$ deg yaw misalignment and one that does implement yaw steering. Their design is based on the work of \cite{kanevDynamicWakeSteering2020} and has further been investigated in \cite{beckerDynamicOpensourceModel2024}. The control framework consists of two parts: (i) the dead-band filter for the wind direction and (ii) the yaw controller.
The dead-band reads in the wind direction measurement $\varphi(k)$ at time step $k$ and updates its wind speed estimate $\hat{\varphi}$ based on the difference between the two values:
\begin{align}
    \hat{\varphi}(k) &= \left\{
    \begin{matrix*}[l]
        \varphi(k) & \text{if } |\varphi(k)- \hat{\varphi}(k-1)|>\varphi_\text{lim}, \\
        \varphi(k) & \text{if } k_i \Delta t\left|\sum_{l=\tau}^{k-1} \varphi(l) - \hat{\varphi}
        (k-1)\right|>\varphi_\text{lim}, \\
        \hat{\varphi}(k-1) & \text{otherwise}.
    \end{matrix*}
    \right.\label{eq:con:deadbandwinddir}
\end{align}
Equation \eqref{eq:con:deadbandwinddir} also consists of a second update law based on the integrated difference between past measurements of $\varphi$ and $\hat{\varphi}$ since the last time step $\tau$ at which $\hat{\varphi}$ was updated. 
The resulting $\hat{\varphi}(k)$ is then used in the LuT, denoted by $f_\text{LuT}$, which returns the yaw set points:
\begin{align}
    \gamma^*(k) =& f_\text{LuT}(\hat{\varphi}(k))\,,\\
    \gamma(k) =& \gamma(k-1) + \text{sign}(\gamma^*(k) - \gamma(k-1)) \cdot \dots \nonumber\\
    &\min\left(\left|\gamma^*(k) - \gamma(k-1)\right|,\; \Delta t \,\delta_\gamma\right)\label{eq:con:yawupdate}
\end{align}
The turbine yaw angle $\gamma(k)$ is then updated based on Equation \eqref{eq:con:yawupdate}: It is either set to $\gamma^*$ if the angle can be reached in $\Delta t$, or changes with the maximum rate of yawing $\delta_\gamma$ for $\Delta t$.
For the baseline controller, we set $\Delta \varphi = 2$ deg and $k_i=0.1$; for the LuT controller, we test $\Delta \varphi \in [2,4]$ deg and $k_i=0.1$. The LuT is calculated with FLORIDyn in steady-state conditions using the same particle swarm optimization as discussed in Section \ref{sec:con:opti}. This way all controllers use the same basis to make decisions. Figure \ref{fig: lut} depicts the LuT for all turbines and wind directions.
A characteristic of this LuT is that downstream turbines show very small misalignment angles to be optimal. This allows the downstream turbine to move slightly out of the way of a partial wake overlap. Toolboxes like FLORIS \citep{nrelFLORISVersion2023} avoid this behavior by setting downstream turbine misalignment angles to $0$~deg by default. Since the CLC controllers tested in this work will also exhibit this behavior, it remains part of the LuT.

\section{Simulation methods}\label{sec: Sim methods}
Section \ref{sec: Results: verification setup} discusses the high fidelity model used as real surrogate, along with the wind direction case. This is followed by Section \ref{sec: Results: tested controllers}, which further specifies the tested controllers and used measurements.

\subsection{``True wind farm'' setup}\label{sec: Results: verification setup}
\begin{figure}
    \centering
    \includegraphics[width=8.5cm]{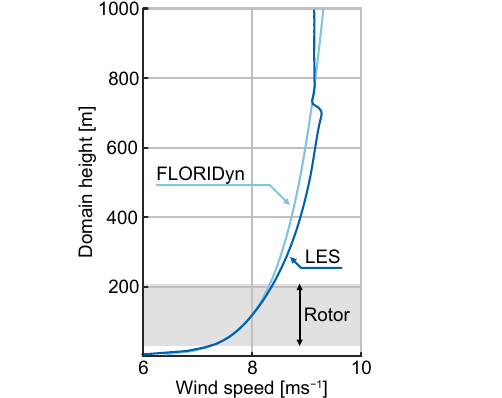}
    \caption{Precursor profile of the wind speed in the precursor and in the FLORIDyn simulation.}
    \label{fig:precursor_shear_ti}
\end{figure}

The experiments are performed in closed-loop with the LES code SOWFA as the true wind farm surrogate \citep{churchfieldNumericalStudyEffects2012}. 
The wind farm is situated in a $5\times 5\times1 \text{ km}$ domain, resolved in $300 \times 300 \times 100$ cells with an average $8 \text{ ms}^{-1}$ free wind speed at hub height and a $0.5$ s time step. All simulations are done with a neutral turbulent precursor developed for $3\cdot10^4\text{ s}$ with a surface roughness length of $2\cdot 10^{-4}\text{ m}$.  This leads to a low turbulence intensity case and more pronounced turbine wakes compared to a high turbulence flow field. 
Based on early flow estimates, the TI in FLORIDyn was set to $5.4\%$, the precursor turbulence intensity at hub height became $I_{0,\text{ u,v,w}} \approx 4~\%$.
The shear resulting from the surface roughness is used to calculate the power-law shear coefficient $\alpha$ based on the mean wind speed magnitude.
The ideal shear coefficient to describe the wind speed across the rotor plane lies between $0.07$ and $0.083$. Based on initial precursor values, $\alpha_s$ was set to $0.071$. 
The mean precursor wind speed magnitude, as well as the FLORIDyn wind speed profile, are depicted in Figure \ref{fig:precursor_shear_ti}~(b). The veer across the rotor plane is less than $2$ deg in the LES and is neglected in FLORIDyn.

\begin{figure}
    \centering
    \includegraphics[width=8.3cm]{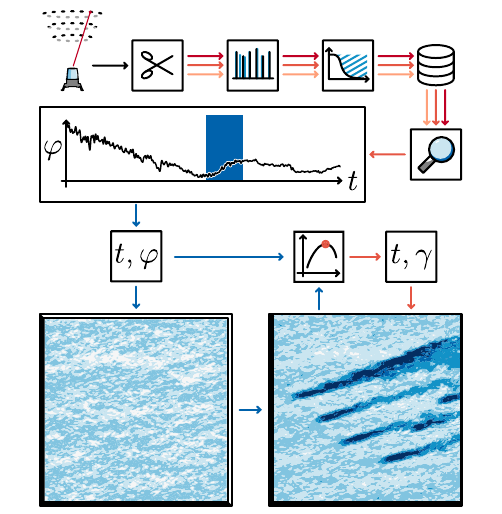}
    \caption{Schematic of the data flow for the verification and test simulations: The initial LiDAR dataset was segmented into time series of sufficient completeness, followed by resampling, interpolation, and zero-phase low-pass filtering. The stored data has then been investigated for time series of long, uninterrupted wind direction trajectories with wake interaction. Out of a 24~h period, one 3~h segment has been chosen to be simulated in a closed loop. The simulation has a precursor with the wind direction changes from the LiDAR data, which is then used to drive the simulation with the wind farm. 
    During the simulation, measurement data is sent from the LES to a controller, which then continuously updates a table with the yaw angle set points for all turbines.}
    \label{fig:data_schematic}
\end{figure}

\begin{figure}
\centering
\includegraphics[width=\scolumnwidth]{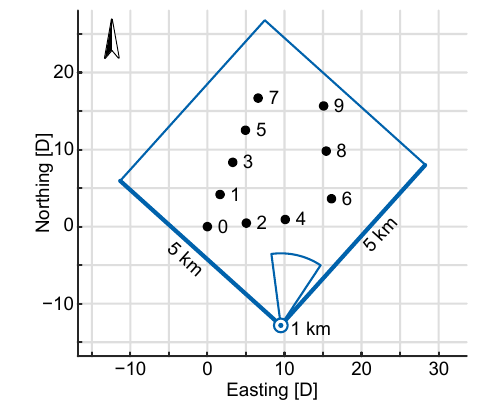}
\caption{Wind farm layout of the investigated case study. The wind farm is oriented as it is in the real world, while the $5\times5\times1$ km domain is rotated to fit the inflow direction. The inflow planes are marked by wider lines. The triangle marks the wind direction range of the case. The wind direction time series is chosen such that initially T6, T8 and T9 interact, and later T0, T1, T3, T5 and T7.}\label{fig:HKN_layout and domain}
\end{figure}

The closed-loop approach is tested in a three-hour long case, which is based on data collected by a vertical LiDAR at the Hollandse Kust Noord (HKN) site on the 28${}^\text{th}$ of March 2023 \citep{knoopWindLidarWind2019}. 
During the measurement campaign (2019 - ongoing), wind speed and direction are recorded at different heights, with values available roughly every $20$ s. First, the data was segmented into parts with sufficient data points. The values at $108$ m and $133$ m were used to resample and interpolate the wind direction at turbine hub height of $119$ m at a regular $20$ s sampling rate. The data was then low-pass filtered with a Butterworth filter with a cut-off frequency of $1/600$ Hz, equivalent to \cite{vandenbroekDynamicWindFarm2024}. The resulting sections were then investigated for completeness and interesting wind direction ramp events. For this work, one sub-section of 3 hours is chosen. Figure \ref{fig:data_schematic} depicts the original data and how it was prepared for the closed-loop tests. 
The wind farm layout and dimensions are depicted in Figure \ref{fig:HKN_layout and domain}, along with the rotated case domain to simulate varying inflow conditions. The same setup has been used to conduct the LuT controller study presented in \cite{beckerDynamicOpensourceModel2024}.

We use OpenFOAM to implement spatial uniform wind direction changes following the predefined time series of wind directions. Within the LES domain, the southwest planes are used as inflow planes, while the northeast planes are outflow planes. 
The farm layout and initial wind direction are rotated to fit the $225$ deg inflow of the turbulent precursor. The rotated layout is centered in the domain to balance the distance to the in- and outflow planes, see Figure \ref{fig:HKN_layout and domain}. The ten DTU 10 MW turbines \citep{bakDTU10MWReference2013} are arranged to copy a subset of the HKN wind farm and are modeled as actuator discs (ADM). 
Turbines modeled as ADMs on a coarser grid tend to overestimate the power generated \citep{martinezComparisonActuatorDisk2012, shapiroFilteredActuatorDisks2019}, which is also an issue with this setup. 
Therefore, with the exception of Figure \ref{fig: Energy 600 s}, we focus on normalized power and energy quantities.

The LES environment is extended by a wind farm-wide controller that receives turbine measurements and can provide set points during the simulation runtime. The measurements received from the LES include quantities like generator power and rotor speed. Wind direction measurements are provided in one of two ways: Either directly from the data that was used to create the precursor or based on probes at hub height at the turbine locations. The former results in a noise-free measurement of the underlying wind direction, while the latter is subject to ambient and turbine-added turbulence.

\subsection{Tested controllers and measurements}\label{sec: Results: tested controllers}
This section specifies the controller configurations tested in tandem with the LES and how they acquire their inputs.

\paragraph*{Model predictive controllers}
All model predictive controllers use the Particle Swarm Optimizer coupled to the 2 degrees-of-freedom baseline function to derive the yaw trajectories; see Section \ref{sec: meth: cost function} - \ref{sec: meth: yaw angle derivation}. Two controllers maximize the energy over $500$ s and $1000$ s. This is done by evaluating the cost function in Equation \eqref{eq:max_energy}. Another controller uses the shifted turbine power signals in time to synchronize control actions with their effect on downstream turbines; see Section \ref{sec: meth: cost function}. This leads to a varying prediction horizon based on the turbines involved in the current optimization problem. Typical values are between $100$ s and $600$ s. The controllers are summarized in Table \ref{tab:closed-loop-controllers}.
\begin{table}
    \centering
    \begin{tabular}{ccccc}
         Controller & Cost function & $\tau_\text{ph}$\\\hline
         PSO, MR& Energy & $0.5\cdot10^4$ s \\
         PSO, MR& Energy & $1\cdot10^4$ s  \\
         PSO, MR& Shifted energy & varying
    \end{tabular}
    \caption{Selection of closed-loop model predictive controllers. PSO refers to the Particle Swarm Optimization, MR to the maximum yawing rate basis function. All controllers have an action horizon of $\tau_\text{ph} = 100$~s and no preview.}
    \label{tab:closed-loop-controllers}
\end{table}
All controllers update the optimal yaw set points every $60$ s and use an action horizon of $100$~s. This allows for a $\pm 30$~deg orientation change based on the maximum yawing rate of $r_\gamma = 0.3~\text{deg s}^{-1}$.

\paragraph*{Reference controllers}
The reference yaw-steering and baseline controllers are based on the dead-band behavior described in Section \ref{sec: meth: reference controllers}. The used threshold and integral gain are given in Table \ref{tab: reference controllers}. In the noise-free environment the controllers are tested with a sampling time of $5$ s to update the set points. With the disturbed wind direction measurements the LuT controllers are updated once every minute based on the past 1-minute average.
\begin{table}
    \centering
    \begin{tabular}{ccccc}
        Steady-state model & $\varphi_\text{lim}$ &  Optimization \\\hline
        FLORIDyn internal & $2$ deg  & PSO offline\\
        FLORIDyn internal & $4$ deg  & PSO offline\\
        Baseline & $2$ deg  & - \\
    \end{tabular}
    \caption{List of reference controller settings. All reference controllers have a $k_i=0.01$~s${}^{-1}$ }
    \label{tab: reference controllers}
\end{table}

\paragraph*{Measurements}
The controllers depend on measurements of wind speed and direction to provide adequate yaw angles. Table \ref{tab:measurement modes} lists the different ways data is provided to the controllers. This concerns the background wind speed $u_\infty$, the wind direction $\varphi$, the sampling time step, and the averaging time of the $\varphi$ measurement.
\begin{table}
    \centering
    \begin{tabular}{c|ccccc}
         & $u_\infty$ & $p$ & $\varphi$& $\Delta t$ & average time $\varphi$\\\hline
        Mode 1 & - & LES & noise free & $15$ s & $0$ s\\
        Mode 2 & - & LES & LES & $15$ s & $60$ s \\
        Mode 3 & given & - & noise free & $5$ s & $0$ s\\
        Mode 4 & given & - & LES & $60$ s & $60$ s
    \end{tabular}
    \caption{Ways for the state estimator and controller to receive data from the simulation. Mode 1 and 2 are relevant to the Ensemble Kalman Filter, Mode 3 and 4 to the reference LuT controllers.}
    \label{tab:measurement modes}
\end{table}
The noise-free data relates to the filtered LiDAR data that was used to drive the precursor. It can, therefore, be considered as an ideal, noise-free signal of the background flow. The same holds for $u_\infty$, which is provided as a constant value to the reference controllers.
The EnKF for the closed-loop controllers integrates new measurements every $15$ s, but a new control decision is taken every $60$ s. The LuT controllers do not have a state and rather act based on the current measurement, hence the lower sampling time.
\begin{figure}
    \centering
    \includegraphics[width=\scolumnwidth]{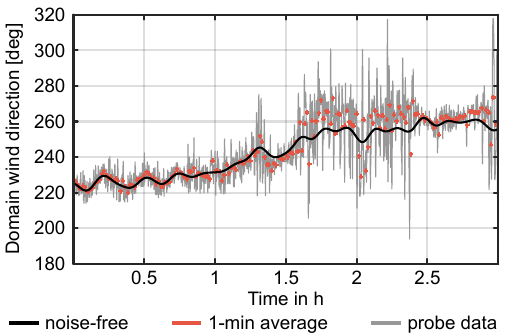}
    \caption{Wind direction data recorded at the location of turbine T7 during the simulation. The source data is based on low-pass filtered field measurements, which is used to drive the LES. The probe data is calculated from the $u$ and $v$ wind speed components at the turbine rotor center. Orange indicates the 1-minute average values of the probe data. The probe data becomes more noisy and biased as the turbine is waked by other turbines, here from $1.5$ to $2.5$ h. Note that the wind direction is given with respect to the LES domain. 
    }
    \label{fig:Probe_background_1min_average}
\end{figure}
Figure~\ref{fig:Probe_background_1min_average} showcases an example of the measurements. The black source data comes from the cleaned and zero-phase low-pass filtered field data; see Figure~\ref{fig:data_schematic}. The grey probe data is recorded in the LES at the rotor center of the turbine, which is, in this example, turbine T7. This was done to mimic a much-simplified version of the measurements a wind vane on a turbine might record. The plot also shows how the noise is reduced by the use of the past 1-minute averaged data instead of the raw probe data. The probe data is characterized by higher noise levels and biases during waked conditions, which poses challenges for the state estimation and the dead-band controllers. Since the wake locations are unique to every simulation, also the LuT controllers have to run online and their control actions can not be pre-computed.

\section{Simulation results}\label{sec: Results: simulation results}
The simulation results first investigate the wake steering behavior in Section \ref{sec: Results: yaw steering}. Section \ref{sec: Results: turbine performance} then compares the performance of the turbines throughout the case study, followed by the farm-wide performance in  Section \ref{sec: Results: farm performance}.

\subsection{Wake steering}\label{sec: Results: yaw steering}
\begin{figure}[h]
    \centering
    \includegraphics[width=\scolumnwidth]{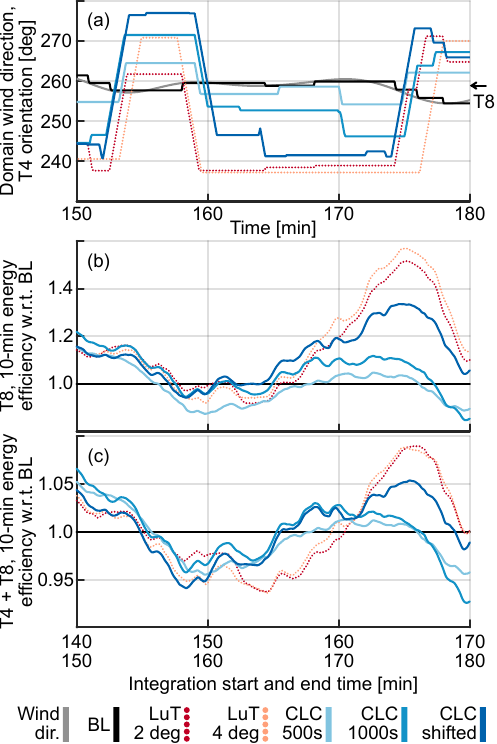}
    \caption{The orientation of turbine T4 over time as a result of the noise-free wind direction measurements is given in (a). The arrow indicates the wind direction in which turbine T8 is located at $10.3$ D distance. The resulting 10-minute energy efficiency of T8 is given in (b), followed by the combined performance of T4 and T8 in (c).}
    \label{fig: turbine orientation perspective}
\end{figure}
Figure~\ref{fig: turbine orientation perspective} (a) depicts the orientation of turbine T4 during the last half hour of the case study. During this time, the wind direction aligns T4 with T8, as indicated by the arrow. All controllers do engage in wake steering to avoid waking turbine T8 at $10.3$ D distance. However, the magnitude of the misalignment differs.
Between minutes $160$ and $175$, the wind direction varies only marginally and does not cross the line between T4 and T8. The LuT controllers engage in the largest misalignment angles, and the Shifted CLC controller acts similarly. The CLC $1000$~s controller exhibits a smaller yaw angle, and lastly, the CLC $500$~s controller shows little-to-no misalignment.
Figure~\ref{fig: turbine orientation perspective} (b) then shows how the 10-minute energy of T8 reacts to the yaw steering efforts of T4: The more aggressive yaw angles by the LuT controllers indeed leads to a better efficiency at T8, increasing its generation by $+56~\%$. The gain of T8, however, comes at the cost of misaligning T4 for a long time. Figure~\ref{fig: turbine orientation perspective} (c) shows the combined efficiency of T4 and T8. The data shows that the period of outperforming the baseline is preceded by a period of underperformance. Based on the shallow misalignment angles, the closed-loop controllers overcome this period earlier than the LuT controllers, confirming the analysis done in Section~\ref{sec: meth: cost function} and that the CLC controllers are working as intended. They do engage in wake steering control and, based on their design, in a more or less aggressive manner. The presented example also shows that the controllers take turbines across a longer distance into account, something with which, e.g. free vortex particle models of the wake can struggle with \citep{vandenbroekDynamicWindFarm2024}.

\subsection{Turbine performance}\label{sec: Results: turbine performance}
\begin{figure}
    \centering
    \includegraphics[width=\scolumnwidth]{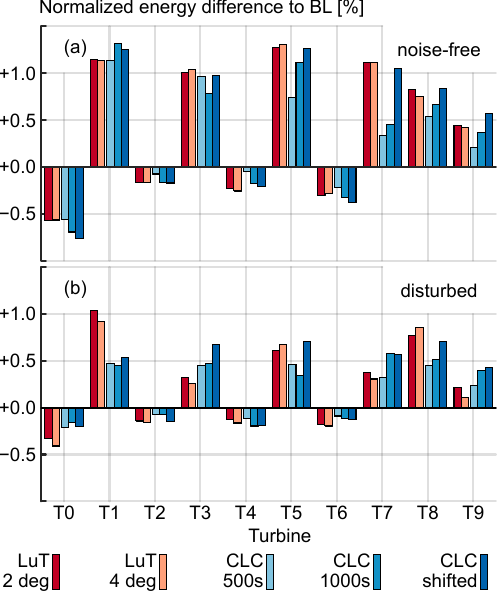}
    \caption{Turbine individual difference between the energy generated in the controlled case and the baseline, normalized by the total baseline energy. The data in (a) relates to the controllers with noise-free wind direction measurements, and the data in (b) to the disturbed measurements. The data is normalized by the respective baseline.}
    \label{fig: bar chart}
\end{figure}

Figure \ref{fig: bar chart} shows the normalized energy gain separated for each turbine. Given the wind direction time series and the farm layout, T0, T2, T4, and T6 are upstream turbines, while T7, T8, and T9 are the most downstream turbines, see Figure \ref{fig:HKN_layout and domain}.

\begin{figure*}[h!] 
    \centering
    \includegraphics[width=\dcolumnwidth]{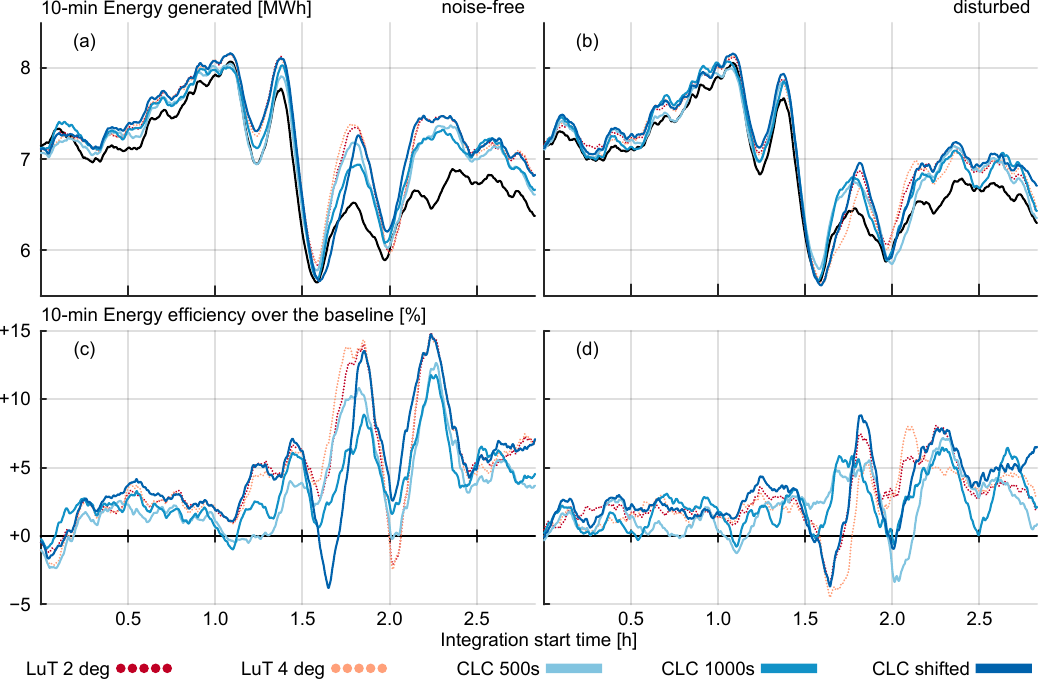}
    \caption{Absolute (a-b) and relative (c-d) energy generated in a sliding time window of 10 minutes. The data in (a) and (c) is based on the noise-free wind direction measurements, and the data in (b) and (d) on the disturbed measurements.}
    \label{fig: Energy 600 s}
\end{figure*}

The data shows that for all controllers, the upstream turbines sacrifice energy by yawing, which is then recouped by the downstream turbines. There are differences in the magnitude: the CLC $500$ s controller amplifies the energy generation of T1, T3, and T5 more than the generation of T7, T8, and T9. It also shows a decreased investment for turbines T0, T2, T4, and T6. This is consistent with the shorted prediction horizon length and with the analysis done in Section \ref{sec: Results: yaw steering}. The CLC $1000$ s shows more committed control actions that further lower the energy generation of the upstream turbines but also result in larger returns for the downstream turbines. The same holds true for the shifted CLC controller, as well as for the LuT controllers. An advantage that the LuT controllers have over the CLC controllers in the noise-free environment is that they are able to engage T5 and T7 consistently.

With disturbed wind direction measurements, the performance of all controllers decreases. The LuT controllers especially sacrifice performance with T3, T5, T7 and T9. This can be a sign that the long-distance wake interactions fail as the wind direction measurement becomes more uncertain. The CLC controllers also sacrifice performance, mainly with T1, T3, and T5.  We attribute this to the lowered yaw investment by T0 due to the more uncertain wind direction.

\subsection{Farm level performance}\label{sec: Results: farm performance}
To investigate the performance of the controllers on a farm level, we compare the wind energy as the power generated over a sliding window integral of ten minutes.
Figure \ref{fig: Energy 600 s} depicts the absolute and relative energy generated by the farm throughout the case study. The absolute data shows the magnitude at which the energy is generated. We note that the largest reductions in energy appear around the $1.5~$h mark, which relates to a brief back-and-forth shift in the wind direction. This crosses the 5-turbine line T0, T1, T3, T5 and T7. During this event, the controllers return their largest gains but also losses. The losses are a product of wake steering as a method: While the turbines are slightly misaligned, the wakes are redirected to one side of the downstream turbine. If the wind direction changes, the wake can only be further redirected up to a certain point at which it is worth redirecting the wake to the other side. During this switch, the baseline wakes have likely already arrived on the new side, which means that the baseline momentarily generates more than the control strategy.

Overall, all tested controllers return consistent gains in the noise-free case, but also in the disturbed case. Here, the performance of the controllers is generally weaker due to the worse sensor data. Qualitatively, the CLC $500$ s controller is the closest to the BL performance and is followed by The CLC $1000$ s controller. The remaining three controllers depict a similar performance.

\begin{figure*}
    \centering
    \includegraphics[width=\dcolumnwidth]{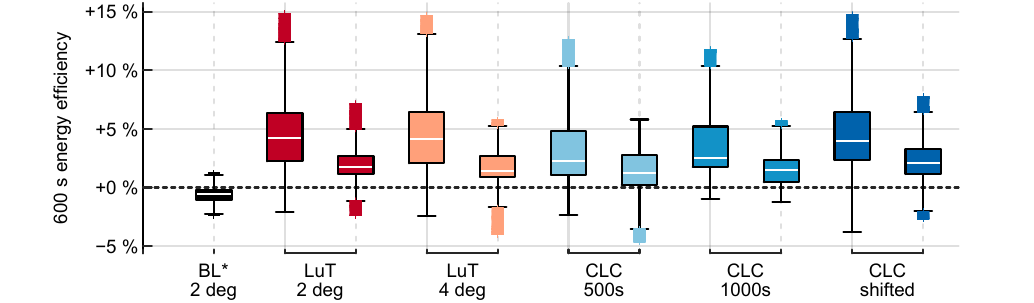}
    \caption{Box plot of the ten-minute wind farm energy efficiency. The data is normalized with the noise-free baseline. Each pair depicts the noise-free performance on the left and the performance with disturbed wind direction measurements on the right. The median values from the left to the right are BL* 2 deg $-0.5$~\%, 
    LuT 2 deg  $+4.2$~\%, LuT* 2 deg  $+1.7$~\%, 
    LuT 4 deg  $+4.1$~\%, LuT* 4 deg  $+1.4$~\%, 
    CLC 500 s $+2.2$~\%, CLC* 500 s $+1.2$~\%, 
    CLC 1000 s $+2.5$~\%, CLC* 1000 s $+1.5$~\%, 
    CLC shifted $+4.0$~\%, CLC* shifted $+2.1$~\%. The disturbed direction simulations are marked with *.}
    \label{fig: box chart}
\end{figure*}

To quantify the performance, the data of Figure \ref{fig: Energy 600 s} (c,d) is summarized as a box plot in Figure \ref{fig: box chart}, with the difference that all data is normalized with the noise-free baseline performance. The two LuT controllers and the time-shifted controller still depict similar performance during noise-free conditions: The LuT 2 deg and 4 deg controllers have a median efficiency of $+4.2$~\% and $+4.1$~\% receptively, the CLC shifted controller has a median of $+4.0$~\%. With disturbed wind direction measurements, the performance drops to a median of $+1.7$~\% and $+1.4$~\% for the LuT 2 deg and 4 deg controllers, while the CLC shifted controller reduces its median to $2.1$~\%. We can conclude that the CLC is, therefore, more robust to sensor noise. This is mainly due to the EnKF, which by design assumes noise to be part of the sensor signal. With noise-free data, this leads to a disadvantage as the measurement is not considered to be fully ``trustworthy'', but for disturbed measurements, this leads to a better estimate. This effect is also visible with the other two CLC controllers: The CLC $500$ s controller drops from a median of $+2.2$~\% to $+1.0$~\%, the CLC $1000$ s controller from $+2.5$~\% to $+1.5$~\%, which is a less significant decrease than the one of the LuT controllers.

A mostly neglected aspect of the cost function is the cost of actuation. It is only captured implicitly as a loss of the power generated by the actuated turbine. 
Actuation costs can include how much and how often the turbines yaw, how much time they spent in misalignment and how their inflow profile looks like. Recent work has suggested ways to create data-driven ways to estimate the loads on a turbine in a surrogate model \citep{liewWindFarmControl2024, guilloreControlorientedLoadSurrogate2024}, but these are not yet included in the FLORIDyn model used in this work. We therefore resort to the yaw travel as quantity of interest.

\begin{figure}
    \centering
    \includegraphics[width=\scolumnwidth]{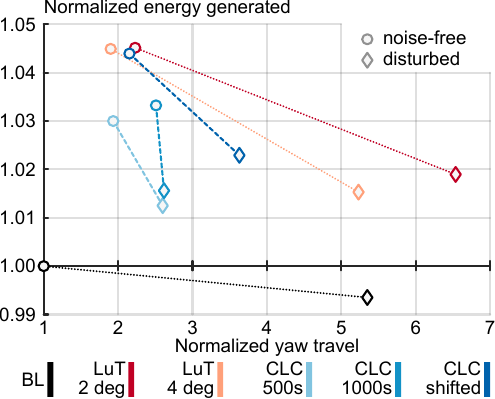}
    \caption{Energy and yaw travel of the six controllers with both noise-free and disturbed wind direction measurements. The data is normalized with the noise-free baseline controller.}
    \label{fig: noise loss}
\end{figure}
Figure \ref{fig: noise loss} shows the overall energy efficiency and normalized yaw travel. A common characteristic of the controllers under noise-free wind direction measurements is that all yaw controllers exhibit an approximately $2$ to $3~\times$ larger amount of yaw travel for a $3$ to $4.5$~\% gain in energy. This ratio changes as the wind direction measurements become disturbed. All controllers are affected by the disturbed measurements, however at a different scale. Similarly to the trend observed with Figure \ref{fig: bar chart}, the LuT are significantly more affected by poor data quality than the CLC controllers: All LuT controllers, including the baseline, exhibit a yaw travel amount that is $\approx\!5$ to $7~\times$ higher than the noise-free baseline, while the CLC controllers remain between $\approx\!2.5$ to $4~\times$ higher values.

\subsection{Computational performance}\label{sec: Results: comp performance}
The closed-loop algorithm has two steps: (i) the state estimation and (ii) the model predictive optimization.
In the case study, the LES is paused for both the state estimation and the optimization. While unrealistic, this allows for more leeway and exploration of the methodology before imposing more challenging conditions.

The state estimation is real-time compatible and roughly requires $6$~s of computational time for simulation,  ensemble combination, and state correction, applied every $15$~s. The optimization, however, is not realtime applicable for larger numbers of turbines. Currently, groups of $\geq 3$ turbines take longer than the allocated $60$~s update time. These numbers are obtained in Matlab for 40 threads.

Future work should investigate the design of a dedicated optimization strategy similar to the serial refine approach \citep{flemingSerialRefineMethodFast2022}.
And to reduce the cost per simulation. In its current form, also the FLORIDyn model used in the optimization uses the spatiotemporal weighting of the OPs. This adds a considerable cost to the model, and while it is necessary for the EnKF, it might not be for the optimization.

\section{Conclusions}\label{sec: Conclusion}
This paper introduces a novel closed-loop control framework to maximize the energy generated by a wind farm under time-varying inflow conditions based on the dynamic wake model FLORIDyn.
The observed case-study results show that the framework can lead up to a median energy gain over 10 minutes of $+4.0$~\% using a shifted cost function and $+2.2$~\% to $+2.5$~\% using an energy maximizing cost function. This, however, falls short behind the tested LuT controllers with gains of $+4.1$~\% and $+4.2$~\%. These results are obtained with noise-free wind direction measurements. If disturbed measurements from the LES are used, the performance of all controllers decreases while the yaw travel increases. However, the closed-loop controllers are less affected by this change, highlighting their robustness due to the Ensemble Kalman Filter and cost-function design. Their median 10-minute energy gain reduces to $+2.1, +1.5,$ and $+1.2$~\%, while the LuT performance decreases to $+1.7$ and $+1.4$~\%.
Future research should explore the scalability of this approach to larger wind farms, as well as in heterogeneous flow conditions.

\section*{Author contributions}
M. B.: conceptualization, methodology, software, validation, investigation, writing -- original draft, visualization.
M.J. v.d.B.:  conceptualization, writing -- review,
D. A.: conceptualization, supervision.
J.W. v.W.: writing -- review \& editing, conceptualization, supervision, resources, funding acquisition.

\section*{Data availability}
The FLORIDyn framework is available with the DOI 10.4121/038777f7-f497-494f-9f61-85d90a00074a . The repository includes the model, the state estimation, and the closed-loop control files.
The simulation data to recreate plots and accompanying information given with the DOI 10.4121/63fa3d6c-6bd2-4efe-9548-ddcaf1bfbeb9
The version of the LES solver SOWFA \citep{churchfieldNumericalStudyEffects2012} used in this work is based on the GitHub repository https://github.com/TUDelft-DataDrivenControl/SOWFA

\section*{Financial disclosure}
This work is part of the research programme ``Robust closed-loop wake steering for large densely spaced wind farms'' with project number 17512, which is (partly) financed by the Dutch Research Council (NWO).

This work has been supported by the SUDOCO project, which receives the funding from the European Union’s Horizon Europe Programme under the grant No. 101122256.

\section*{Conflict of interest}

The authors declare no potential conflict of interest.

\bibliography{the_full_bib}



\appendix
\section{Parameter tuning}\label{app: parameters}
The total number of parameters is high in the proposed framework. Table \ref{tab:parameters} lists all parameters and constants: Six parameters to parametrize the OP weighting, one to limit the advection speed, eleven for the wake and turbine model, eight for the state estimation, and six for the optimization. Ideally, one would tune all parameters to fit a generalized scenario. The number of parameters to tune is high, which makes it difficult to replicate the related work \citep{howlandOptimalClosedloopWake2020, doekemeijerClosedloopModelbasedWind2020, digheSensitivityAnalysisBayesian2022}. We, therefore, resort to Latin Hypercube sampling of the remaining parameters and test the different combinations in a $20$-minute sub-set simulation of the final setup.
For each simulation, a set of farm-wide error quantities is calculated:
\begin{itemize}
    \item Mean bias in turbine power
    \item Mean absolute turbine power error
    \item Mean squared turbine power error
    \item Mean squared farm power error
    \item Farm power bias
    \item Mean squared turbine power error weighted by the predicted spread of the EnKF
\end{itemize}
This is complemented by turbine individual error quantities:
\begin{itemize}
    \item Best possible power correlation based on a variable signal time-shift
    \item Time-shift at which the best correlation is achieved
    \item Turbine power bias
    \item Absolute power error
    \item EnKF weighted power error
\end{itemize}
Out of $600$ simulations, the parameter combinations that perform the best are compared for similarities. The remaining parameters are tuned manually, informed by parameters that were deemed ideal in previous studies.
During this process, $\eta=1$ was locked to further reduce the number of parameters. However, the turbine model used in the simulations overestimates the power, see Section \ref{sec: Results: verification setup}. Since the power was used to correct the wind speed, the state estimation returns a $\approx 8~\%$ too higher wind speed. This was captured by the error quantities and corrected by adjusting the free parameters. The wake advection factor $d$ is smaller, and the remaining parameters lead to a faster wake recovery. 
The resulting parameter set may work for this setup but may not be ideal for generalization.
\begin{table*}
    \centering
    \begin{tabular}{cccccc}
         Parameter & Use & Range & Selected & Unit & Framework component \\\hline
         $\sigma_{\text{w,dw},\varphi}$ & weighting OP, Dir. & - & $2.87$ & D & FLORIDyn \\
         $\sigma_{\text{w,cw},\varphi}$ & weighting OP, Dir. & - & $2.87$ & D & FLORIDyn \\
         $\sigma_{\text{w,t},\varphi}$  & weighting OP, Dir. & - & $50$  & s & FLORIDyn\\
         $\sigma_{\text{w,dw},u}$ & weighting OP, Vel. & $[0.3,\,3.5]$ & $0.6966$  & D & FLORIDyn \\
         $\sigma_{\text{w,cw},u}$ & weighting OP, Vel. & $[0.3,\,2]$ & $0.3570$ &  D & FLORIDyn\\
         $\sigma_{\text{w,t},u}$  & weighting OP, Vel. & $[100,\,300]$ & $206.2331$ &  s & FLORIDyn\\
         $d$ & Advection factor & $[0.5,\,1.0]$ & $0.7396$  & - & FLORIDyn \\\hline
         $\eta$ & Turbine efficiency & - &  $1.0$   & - & Turbine model \\
         $p_p$ & Yaw exponent (Power) & $[1.7,\, 2.7]$ &  $2.2$ &   - & Turbine model\\
         $\alpha$ & Near wake length & - &  $2.32$  &  - & Wake model\\
         $\beta$ & Near wake length & $[0.07,\,0.39]$ &  $0.154$  &  - & Wake model\\
         $k_a$& Wake expansion & $[0.17,\,0.92]$ &  $0.38371$  &  - & Wake model \\
         $k_b$& Wake expansion & - &  $0.003678$   & - & Wake model\\
         $k_{fa}$& Added turbulence & - &  $0.73$  & - & Wake model\\
         $k_{fb}$& Added turbulence & $[0,\,8]$ &  $0.8325$  & - & Wake model \\
         $k_{fc}$& Added turbulence & $[0,\,0.5]$ &  $0.0325$   & - & Wake model \\
         $k_{fd}$& Added turbulence & - &  $-0.32$  & - & Wake model  \\
         $k_\text{TI}$& TI spread & $[1,\,4]$ &  $3$  & - & Wake model \\\hline
         $l_{\text{loc.},\varphi}$& Localisation, Dir. & - &  $2.8$ & D & EnKF \\
         $\sigma_{\mu,\varphi}$ & Process noise, Dir. & - &  $0$*   & deg & EnKF  \\
         $\sigma_{\nu, \varphi}$ & Measurement noise, Dir. & - &  $3$  & deg & EnKF \\
         $l_{\text{loc.},u}$& Localisation, Vel. & $[3.5,\,8]$ &  $6.8011$   & D & EnKF \\
         $\sigma_{\mu, u}$ & Process noise, Vel. & $[0.1,\,0.5]$ &  $0.1991$   & m/s & EnKF \\
         $\sigma_{\nu, p}$ & Measurement noise, Pow. & $[0.01,\,0.3]$ &  $0.08$   & MW & EnKF \\
         $n_\text{e}$ & Number of ensembles & - &  $50$  & -  & EnKF\\
         $k_\text{enkf}$ & Estimation sample time & - &  $3$  & $\Delta t$ & EnKF\\\hline
         $\tau_\text{ah}$ & Action horizon & - &  $20$   & $\Delta t$ & eMPC \\
         $r_{\gamma}$ & Yaw rate limit & - &  $0.3$   & $\text{deg s}^{-1}$ & eMPC \\
         $n_\text{iter, max}$ & Max. optimization iterations & - &  $20$  & - & eMPC \\
         $k_\text{mpc}$ & Optimization sample time & - &  $12$   & $\Delta t$ & eMPC \\
         $\gamma_\text{max}$ & Yaw limitation & - &  $33$   & $\text{deg}$ & eMPC \\
         $\gamma_\text{min}$ & Yaw limitation & - &  $-33$   & $\text{deg}$ & eMPC \\
    \end{tabular}
    \caption{Collection of all parameters and constants, their use, the investigated range, and selected value, as well as the component they belong to. The turbine diameter $\text{D}=178.4\,\text{m}$ and time step $\Delta t = 5$~s are used to normalize some parameters. Parameters without range have not been tuned. *The process noise was unintentionally set to $\sigma_{\mu,\varphi}=0$~deg, and should, for future experiments, be set to a higher value.}
    \label{tab:parameters}
\end{table*}

\end{document}